\documentclass{article}
\usepackage[utf8]{inputenc}
\usepackage[a4paper, left=25mm, right=25mm, top=20mm, bottom=20mm]{geometry}
\usepackage{mathtools}
\usepackage{physics}
\usepackage{csquotes}
\usepackage[sorting=none, style=nature]{biblatex} 
\usepackage{amsmath}
\DeclareMathOperator{\arcsinh}{arcsinh}
\usepackage{setspace}
\usepackage{algorithm}
\usepackage{algpseudocode}
\usepackage{amsfonts}
\usepackage{scrextend}
\usepackage{verbatim}

\usepackage{authblk}
\newcommand{\indep}{\perp \!\!\! \perp}

\usepackage{float} 

\title{Classical modelling of a lossy Gaussian bosonic sampler}
\author[1]{M.V. Umanskii}
\author[1,2]{A.N. Rubtsov}
\affil[1]{Department of Physics, Lomonosov Moscow State University, Leninskie gory 1, Moscow 119991, Russia}
\affil[2]{Russian Quantum Center, Bolshoy Bulvar 30, bld. 1, Skolkovo, Moscow 121205, Russia}
\date{\today}                     %% if you don't need date to appear
\begin{document}

\maketitle

%\section{Abstract}

\abstract{ Gaussian boson sampling (GBS) is considered a candidate problem for demonstrating quantum advantage. We propose an algorithm for approximate classical simulation of a lossy GBS instance. The algorithm relies on the Taylor series expansion, and increasing the number of terms of the expansion that are used in the calculation yields greater accuracy. The complexity of the algorithm is polynomial in the number of modes given the number of terms is fixed. We describe conditions for the input state squeezing parameter and loss level that provide the best efficiency for this algorithm (by efficient we mean that the Taylor series converges quickly). In recent experiments that claim to have demonstrated quantum advantage, these conditions are satisfied; thus, this algorithm can be used to classically simulate these experiments. 
    }

\section{Introduction}

Quantum computers are computational devices which operate using phenomena described by quantum mechanics. Therefore, they can carry out the operations which are not available for classical computers. The ability of a quantum computer to solve a specific task faster that any classical computer is usually referred to as quantum advantage. Although quantum algorithms that provide exponential speedup over classical ones are known, they are hard to implement in practice. Examples of such algorithms include Shor's algorithm of factoring integers\cite{Shor_1997}, that works in polynomial time, whereas all classical algorithms require exponential time. Modern quantum computers are far from experimentally demonstrating quantum advantage on basic problems like integer factorization.

Boson sampling\cite{Lund_2014} is a problem that was proposed as a good candidate for demonstrating quantum advantage due to its nature. A boson sampler is a linear-optical device that consists of non-classical sources of indistinguishable photons, a multichannel interferometer mixing photons of different sources, and photon detectors at the output channels of the interferometer. In the original proposal, the indistinguishable photons were prepared in Fock states. The problem then is to calculate the photon statistics after the interferometer, given an input state and the interferometer matrix. The relevant parameters are the number of modes $N$ and the total number of photons injected in the interferometer $M$. Experimentally it corresponds to performing multiple measurements of the photon counts at the outputs of such a device\cite{Gard_2015}.

Due to the technological complexity of generating Fock states, several variants of the original boson sampling problem have been proposed. They aim at improving the photon generation efficiency and increasing the scale of implementations. One such example is the Scattershot boson sampling, which uses many parametric down-conversion sources to improve the single photon generation rate. It has been implemented experimentally using a 13-mode integrated photonic chip and six PDC photon sources\cite{SBS_exp}. 

Another variant is the Gaussian boson sampling\cite{Hamilton_2017}\cite{PhysRevLett.113.100502}, in which Gaussian states are injected into the interferometer instead of Fock states. Gaussian input states can be generated using PDC sources, and it allows the non-classical input states to be prepared deterministically. In this variant, the relative input photon phases can affect the sampling distribution. Experiments were carried out with $N=12$\cite{Zhong_2019}, $N=100$\cite{Zhong_2020} and $N=144$\cite{Zhong_2021}. The latter implementations used PPKTP crystals as PDC sources and employs an active phase locking mechanism to ensure a coherent superposition.

Any experimental set-up, of course, differs from the idealized model considered in theoretical modelling. Bosonic samplers suffer from two fundamental types of imperfections. First, the parameters of a real device, such as the reflection coefficients of the beam splitters and the phase rotations, are never known exactly. A small change in the interferometer parameters can affect the sampling statistics drastically, so that modelling of an ideal device no longer makes much sense. Another type of imperfections is photon losses. These losses happen because of imperfections in photon preparation, absorption inside the interferometer and imperfect detectors and coupling. 

There are different ways of modelling losses, for example by introducing extra beam splitters\cite{Oh_2021} or replacing the interferometer matrix  by a combination of lossless linear optics transformations and the diagonal matrix that contains transmission coefficients\cite{Garc_a_Patr_n_2019}. In the algorithm described in this paper we will assume that losses occur on the inputs of the interferometer, and we will describe the exact way that we model them.

Imperfections in  middle-sized systems make them, in general, easier to emulate with classical computers\cite{https://doi.org/10.48550/arxiv.2106.01445}. It was shown\cite{Qi_2020} that with the increase of losses in a system the complexity of the task decreases. When the number of photons $M'$ that arrive at the outputs is less than $\sqrt{M}$, the problem of boson sampling can be efficiently solved using classical computers. On the other hand, if the losses are low, the problem remains hard for classical computers\cite{Aaronson_2016}.

In this paper we propose a classical algorithm for calculating probabilities of output states in a GBS problem. The algorithm uses Taylor series expansion, and it converges faster depending on the parameters of the problem, namely the amount of losses in the system and the squeezing parameter of the input states. The higher the losses in the system, the less orders of the series are needed to approximate the probability of observing a given output state.

The work by Oh et al.\cite{oh2023classical} used the following approach to simulating GBS: the covariance matrix of the output Gaussian state was decomposed into "quantum"\ and "classical"\ parts, and the "quantum"\ part was simulated using matrix product states and the "classical"\ part was simulated by random displacement. Thus, when the photon loss rate is high, computational complexity of this algorithm is reduced.

The algorithm that we propose in this paper uses some similar ideas, namely the zeroth order of the Taylor series may be considered the "classical"\ part that is computed quite easily, while the remaining terms are the "quantum"\ part that is more computationally complex. The contribution of this "quantum"\ part is smaller when the losses in the system are high; thus, our algorithm also has optimal conditions that depend on the magnitude of losses. We also analyze some recent GBS implementations to compare the conditions in those experiments with the optimal conditions for our algorithm.    

\section{Problem specification}

Let us first consider a lossless linear-optics interferometer with a transmission matrix $U$:

\begin{equation}
    \hat{a}_i^\dagger = \sum_j U_{ij} \hat{d}_j^\dagger,~~~ \hat{a}_i = \sum_j U_{ij}^*\hat{d}_j
\end{equation}
where creation operators acting on the $i$-th input and output modes are denoted $a_i^\dagger$ and $d_i^\dagger$. Suppose the input modes are injected with single-mode squeezed states:

\begin{equation}
    \ket{\psi} = e^{\sum_i \frac{\alpha_i}{2} (\hat{a}_i^\dagger)^2} \ket{0}.
\end{equation}

The goal is to calculate the probability of detecting $n_1$ photons in the first output mode, $n_2$ photons in the second output mode and so on. This probability can be calculated in the following way:

\begin{equation}
    Tr\left\{\hat{\rho}_{out} \hat{\Vec{n} }\right\} = Tr\left\{\hat{\rho}_{out} \bigotimes_i \ket{n_i}\bra{n_i}\right\},
\end{equation}
where $\hat{\rho}_{out}$ is the density matrix of the output state.

\subsection{Modelling losses}

In real-life bosonic samplers there will always be losses. Here we will model them by substituting

\begin{equation}
    a_i^\dagger \longrightarrow c a_i^\dagger + s b_i^\dagger,
\end{equation}
where $b_i^\dagger$ acts on a mode that we cannot observe, and $c^2+s^2=1$, $c, s \in \mathbb{R}$. Now, the goal is to compute the same probability ($Tr\left\{\hat{\rho_{out}} \hat{\Vec{n} }\right\}$), but taking losses into account. Input state will now be

\begin{equation}
    \ket{\psi'} = e^{\sum_i \frac{\alpha_i}{2} (c \hat{a}_i^\dagger + s \hat{d}_i^\dagger)^2} \ket{0_a 0_b},
\end{equation}
and we now take partial trace over all loss modes when calculating the density matrix:

\begin{equation}
    \hat{\rho} = Tr_b \left\{ e^{\sum_i \frac{\alpha_i}{2} (c \hat{a}_i^\dagger + s \hat{d}_i^\dagger)^2} \ket{0_a 0_b} \bra{0_a0_b} e^{\sum_i \frac{\alpha_i}{2} (c \hat{a}_i^\dagger + s \hat{d}_i^\dagger)^2}\right\}.
\end{equation}

\section{Algorithm derivation}

Let us consider a single mode:

\begin{equation}
    \ket{\psi'} = e^{\frac{\alpha}{2} (c \hat{a}^\dagger + s \hat{b}^\dagger)^2} \ket{0_a 0_b},
\end{equation}
\begin{equation}
    \hat{\rho} = Tr_b \left\{ e^{\frac{\alpha}{2}(c \hat{a}^\dagger + s \hat{b}^\dagger)^2} \ket{0_a 0_b} \bra{0_a0_b} e^{\frac{\alpha}{2}(c \hat{a} + s \hat{b})^2} \right\}.
\end{equation}

\subsection{Calculating partial trace}

We start by applying the Hubbard–Stratonovich transformation\cite{PhysRevLett.3.77}\cite{1957SPhD....2..416S}

\begin{equation}
    e^{\frac{\hat{A}^2}{2}} = \frac{1}{\sqrt{2 \pi}} \int e^{\xi \hat{A} - \frac{\xi^2}{2}} d \xi
\end{equation}
to both exponents in the density matrix operator. This gives us the following:

\begin{equation*}
    \hat{\rho} = \frac{1}{2 \pi} \int Tr_b \left\{ e^{\xi \sqrt{\alpha} (c \hat{a}^\dagger + s \hat{b}^\dagger)} \ket{0_a 0_b} \bra{0_a 0_b} e^{\Tilde{\xi} \sqrt{\alpha} (c \hat{a} + s \hat{b})} \right\} e^{-\frac{\xi^2 + \Tilde{\xi}^2}{2}} d \xi d \Tilde{\xi} = 
\end{equation*}
\begin{equation*}
    = \frac{1}{2 \pi \alpha } \int Tr_b \left\{ e^{\xi \sqrt{\alpha} (c \hat{a}^\dagger + s \hat{b}^\dagger)} \ket{0_a 0_b} \bra{0_a 0_b} e^{\Tilde{\xi} \sqrt{\alpha} (c \hat{a} + s \hat{b})} \right\} e^{-\frac{(\xi \sqrt{\alpha})^2 + (\Tilde{\xi} \sqrt{\alpha})^2}{2 \alpha}} d (\xi \sqrt{\alpha}) d (\Tilde{\xi} \sqrt{\alpha}).
\end{equation*}

Let us redefine $\xi \sqrt{\alpha} \longrightarrow \xi$, $\Tilde{\xi} \sqrt{\alpha} \longrightarrow \Tilde{\xi}$ for convenience:

\begin{equation}
    \hat{\rho} = \frac{1}{2 \pi \alpha} \int Tr_b \left\{ e^{\xi  (c \hat{a}^\dagger + s \hat{b}^\dagger)} \ket{0_a 0_b} \bra{0_a 0_b} e^{\Tilde{\xi}  (c \hat{a} + s \hat{b})} \right\} e^{-\frac{\xi^2 + \Tilde{\xi}^2}{2 \alpha}} d \xi d \Tilde{\xi}.
\end{equation}

We can now calculate the partial trace over loss modes:

\begin{equation*}
    Tr_b \left\{ e^{\xi (c \hat{a}^\dagger + s \hat{b}^\dagger)} \ket{0_a 0_b} \bra{0_a 0_b} e^{\Tilde{\xi} (c \hat{a} + s \hat{b})} \right\} = 
\end{equation*}
\begin{equation*}
    = e^{\xi c \hat{a}^\dagger} \ket{0_a} \bra{0_a} e^{\Tilde{\xi} c \hat{a}} \cdot Tr \left\{  e^{\xi s \hat{b}^\dagger} \ket{0_b} \bra{0_b} e^{\Tilde{\xi} s \hat{b}} \right\}  = 
\end{equation*}
\begin{equation*}
    = e^{\xi c \hat{a}^\dagger} \ket{0_a} \bra{0_a}  e^{\Tilde{\xi} c \hat{a}} \cdot \bra{0_b} e^{\Tilde{\xi} s \hat{b}} e^{\xi s \hat{b}^\dagger} \ket{0_b} .
\end{equation*}

The following expression can be simplified:

\begin{equation*}
    \bra{0_b} e^{\Tilde{\xi} s \hat{b}}  e^{\xi s \hat{b}^\dagger} \ket{0_b} =
\end{equation*}
\begin{equation*}
 = \bra{0_b} (1+\Tilde{\xi} s \hat{b} + \frac{1}{2} (\xi s \hat{b})^2 + ...)(1+\xi s \hat{b^\dagger} + \frac{1}{2} (\xi s \hat{b^\dagger})^2 + ...) \ket{0_b} =
\end{equation*}
\begin{equation*}
 = \left( \bra{0_b}+\Tilde{\xi} s \bra{1_b} + \frac{1}{\sqrt{2}} (\Tilde{\xi} s)^2 \bra{2_b} + ...\right) \left( \ket{0_b}+\xi s \ket{1_b} + \frac{1}{\sqrt{2}} (\xi s)^2 \ket{2_b} + ... \right)   = 
\end{equation*}
\begin{equation*}
 = 1 + \xi \Tilde{\xi} s^2 + \frac{1}{2} (\xi \Tilde{\xi} s^2)^2 + ... = e^{\xi \Tilde{\xi} s^2}.
\end{equation*}

The density matrix now can be written in the following way:

\begin{equation}
    \hat{\rho} = \frac{1}{2 \pi \alpha} \int e^{\xi c \hat{a}^\dagger} \ket{0} \bra{0} e^{\Tilde{\xi} c \hat{a}} \cdot e^{-\frac{\xi^2 + \Tilde{\xi}^2}{2 \alpha} + \xi \Tilde{\xi} s^2} d \xi d \Tilde{\xi}.
\end{equation}

\subsection{Switching between probability density functions}

We can view this integral as taking an expected value over a $2$-dimensional normal distribution. $\xi$ and $\Tilde{\xi}$ then become normally distributed random variables with mean vector equal to zero. Their covariance matrix has the following form:

\begin{equation}
    \Sigma = \begin{pmatrix}
    1/\alpha & -s^2  \\
    -s^2 & 1/\alpha \\
    \end{pmatrix}^{-1} = \frac{1}{1/\alpha^2-s^4}\begin{pmatrix}
    1/\alpha & s^2  \\
    s^2 & 1/\alpha \\
    \end{pmatrix}.
\end{equation}

Then we can write

\begin{equation}
   \hat{\rho} = \frac{(det \Sigma)^{1/2}}{\alpha} \frac{1}{2 \pi (det \Sigma)^{1/2}} \int e^{\xi c \hat{a}^\dagger} \ket{0} \bra{0} e^{\Tilde{\xi} c \hat{a}} e^{-\frac{\xi^2 + \Tilde{\xi}^2}{2 \alpha} + \xi \Tilde{\xi} s^2} d \xi d \Tilde{\xi} = 
\end{equation}
\begin{equation}
    = \frac{(det \Sigma)^{1/2}}{\alpha} \cdot \mathbb{E}_{\mathbb{N}(0, \Sigma)} \left[e^{\xi c \hat{a}^\dagger} \ket{0} \bra{0} e^{\Tilde{\xi} c \hat{a}}\right],
\end{equation}
where $\mathbb{E}_{\mathbb{N}(0, \Sigma)}$ denotes averaging over the $2$-dimensional normal distribution $\mathbb{N}(0, \Sigma)$.

The expression $e^{\xi c \hat{a}^\dagger} \ket{0} \bra{0} e^{\Tilde{\xi} c \hat{a}}$ is troublesome to calculate, since there are two different variables $\xi$ and $\Tilde{\xi}$. We want to arrive somehow at an expression with only one such variable, i.e. $e^{\xi c \hat{a}^\dagger} \ket{0} \bra{0} e^{\xi c \hat{a}}$, which we will denote $\hat{\nu}(\xi c)$.

We now will choose normally distributed random variables $\xi_0, \chi, \Tilde{\chi} \in \mathbb{R}$ such that $\xi = \xi_0 + \chi, ~ \Tilde{\xi} = \xi_0 + \Tilde{\chi}$ and the distributions over $\xi, \Tilde{\xi}$ and $\xi_0, \chi, \Tilde{\chi}$ have the same moments:

\begin{equation}
    \begin{cases}
        \overline{\xi^2}=\overline{(\xi_0 + \chi)^2} = \overline{\xi_0^2}+2\overline{\xi_0 \chi} + \overline{\chi^2}, \\
        \overline{\Tilde{\xi}^2}=\overline{(\xi_0 + \Tilde{\chi})^2} = \overline{\xi_0^2}+2\overline{\xi_0 \Tilde{\chi}} + \overline{\Tilde{\chi}^2}, \\
        \overline{\xi \Tilde{\xi}} = \overline{(\xi_0+\chi)(\xi_0+\Tilde{\chi})} = \overline{\xi_0^2}+\overline{\xi_0 \chi}+\overline{\xi_0 \Tilde{\chi}} + \overline{\chi \Tilde{\chi}}.
    \end{cases}
\end{equation}

We have some freedom in choosing these variables; we will set $\overline{\xi_0 \chi}=\overline{\xi_0 \Tilde{\chi}}=0$ so that $\xi_0 \indep \chi$ and $\xi_0 \indep \Tilde{\chi}$. Then the covariance matrix $\Gamma$ of $\xi_0, \chi, \Tilde{\chi}$ will be determined by one parameter $h = \overline{\chi \Tilde{\chi}}$:

\begin{equation}
    \begin{cases}
        \overline{\xi^2}= \overline{\xi_0^2} + \overline{\chi^2}, \\
        \overline{\Tilde{\xi}^2} = \overline{\xi_0^2} + \overline{\Tilde{\chi}^2}, \\
        \overline{\xi \Tilde{\xi}} = \overline{\xi_0^2} + h.
    \end{cases}
\end{equation}

\begin{equation}
    \begin{cases}
        \overline{\xi_0^2} = \overline{\xi \Tilde{\xi}} - h = \frac{s^2}{1/\alpha^2 - s^4} - h, \\
        \overline{\chi^2} = \overline{\Tilde{\chi}^2} = \overline{\xi^2}-\overline{\xi \Tilde{\xi}} + h = \frac{1/\alpha - s^2}{1/\alpha^2 - s^4} + h = \frac{1}{1/\alpha + s^2} + h.
    \end{cases}
\end{equation}

Note that $-\frac{1}{1/\alpha + s^2} \leq h \leq \frac{s^2}{1/\alpha^2 - s^4}$. We will later find an optimal way to choose $h$. The density matrix in terms of the new variables $\xi_0, \chi, \Tilde{\chi} \in \mathbb{N}(0, \Gamma)$ is

\begin{equation}
    \hat{\rho}= \frac{(det \Sigma)^{1/2}}{\alpha} \cdot \mathbb{E}_{\mathbb{N}(0, \Gamma)} \left[e^{(\xi_0 + \chi) c \hat{a}^\dagger} \ket{0} \bra{0} e^{(\xi_0 + \Tilde{\chi}) c \hat{a}}\right].
\end{equation}

\subsection{Taylor series expansion}

We now consider the Taylor series of the expression $e^{(\xi_0 + \chi) c \hat{a}^\dagger} \ket{0} \bra{0} e^{(\xi_0 + \Tilde{\chi}) c \hat{a}}$, leaving only $\xi_0$ in the exponent:

\begin{equation*}
e^{(\xi_0+\chi) c \hat{a}^\dagger} \ket{0} \bra{0} e^{(\xi_0+\Tilde{\chi}) c \hat{a}} =  e^{\chi c \hat{a}^\dagger} e^{\xi_0 c \hat{a}^\dagger} \ket{0} \bra{0} e^{\xi_0 c \hat{a}} e^{\Tilde{\chi} c \hat{a}} = 
\end{equation*}
\begin{equation*}
= e^{\chi c \hat{a}^\dagger} \hat{\nu}(\xi_0 c) e^{\Tilde{\chi} c \hat{a}} = \left(1 + \chi c \hat{a}^\dagger + \frac{(\chi c \hat{a}^\dagger)^2}{2} + ... \right) \hat{\nu}(\xi_0 c) \left(1 + \Tilde{\chi} c \hat{a} + \frac{(\Tilde{\chi} c \hat{a})^2}{2} + ... \right).
\end{equation*}

Each term in the expression will be proportional to 

\begin{equation*}
    \chi^n \Tilde{\chi}^m \cdot (\hat{a}^\dagger )^n \hat{\nu}(\xi_0 c) \hat{a}^m,
\end{equation*}
and since $\xi_0 \indep \chi$ and $\xi_0 \indep \Tilde{\chi}$, the integral over $\xi_0, \chi, \Tilde{\chi}$ can be written as a product of integrals over $\xi_0$ and $\chi, \Tilde{\chi}$. The latter can be taken analytically:

\begin{equation}
    \int \chi^n \Tilde{\chi}^m d\chi d\Tilde{\chi} \propto \mathbb{E}_{\mathbb{N}(0, \Gamma)} \left[ \chi^n \Tilde{\chi}^m \right],
\end{equation}
which can be calculated using Wick's probability theorem.

\subsection{Choosing $\Gamma$}

The idea consists in minimizing the "perturbation parameter"\ so that each subsequent order of the Taylor series expansion has less impact on the expression. Since higher orders of the expansion contain higher powers of $c^2$ and higher moments $\mathbb{E}_{\mathbb{N}(0, \Gamma)} \left[ \chi^n \Tilde{\chi}^m \right]$, and these moments can be calculated via second moments $\overline{\chi^2}=\overline{\Tilde{\chi}^2}$ and $\overline{\chi \Tilde{\chi}}=h$, the role of the "perturbation parameter" is played by $\varepsilon = c^2 \cdot \max(\overline{\chi^2}, |\overline{\chi \Tilde{\chi}}|)$.

Let us consider the conditions that must be satisfied by $h$. Firstly, $h$ must satisfy $-\frac{1}{1/\alpha + s^2} \leq h \leq \frac{s^2}{1/\alpha^2 - s^4}$, because $\overline{\xi_0^2} \geq 0$ and $\overline{\chi^2} \geq 0$. Secondly, since $\Gamma$ is a covariance matrix, its eigenvalues must be non-negative. The eigenvalues of $\Gamma$ are $\overline{\xi_0^2}$, $\overline{\chi^2}-h$ and $\overline{\chi^2}+h$. Thus, $h$ needs to satisfy

\begin{equation}
    \overline{\chi^2}+h \geq 0 \iff \frac{1}{1/\alpha + s^2} + 2h \geq 0 \iff h \geq -\frac{1}{2}\frac{1}{1/\alpha + s^2}.
\end{equation}

The minimum of $\max(\overline{\chi^2}, |h|)$ is realised when $h=-\overline{\chi^2}=-\frac{1}{2}\frac{1}{1/\alpha + s^2}$.

\subsection{Multimode case}

Let's apply the steps described above to the case of $N$ modes. We start with an input state

\begin{equation*}
    \ket{\psi'^{(N)}} = \prod_{i=1}^{N}e ^{\frac{\alpha}{2} (c \hat{a}_i^\dagger + s \hat{b}_i^\dagger)^2} \ket{0_a 0_b}.
\end{equation*}

We construct a density matrix and take the partial trace over loss modes:

\begin{equation*}
\hat{\rho} = Tr_b \left\{ e^{\sum_{i} \frac{\alpha}{2} (c \hat{a}_i^\dagger + s \hat{b}_i^\dagger)^2} \ket{0_a 0_b} \bra{0_a 0_b} e^{\sum_{i} \frac{\alpha}{2} (c \hat{a}_i + s \hat{b}_i)^2} \right\}.
\end{equation*}

We apply the Hubbard–Stratonovich transformation $2N$ times, resulting in an integral over $\prod_{i=1}^N d \xi_i d \Tilde{\xi}_i$:

\begin{equation*}
    \hat{\rho} = \frac{1}{(2 \pi)^N} \int Tr_b \left\{ e^{\sum_i \xi_i \sqrt{\alpha} (c \hat{a}_i^\dagger + s \hat{b}_i^\dagger)} \ket{0_a 0_b} \bra{0_a 0_b} e^{\sum_i \Tilde{\xi}_i \sqrt{\alpha} (c \hat{a}_i + s \hat{b}_i)} \right\}  e^{-\sum_i \frac{\xi_i^2 + \Tilde{\xi}_i^2}{2}} \prod_i d \xi_i d \Tilde{\xi}_i = 
\end{equation*}
\begin{equation*}
    = \frac{1}{(2 \pi \alpha)^N} \int Tr_b \left\{ e^{\sum_i \xi_i \sqrt{\alpha} (c \hat{a}_i^\dagger + s \hat{b}_i^\dagger)} \ket{0_a 0_b} \bra{0_a 0_b} e^{\sum_i \Tilde{\xi}_i \sqrt{\alpha} (c \hat{a}_i + s \hat{b}_i)} \right\} \cdot
\end{equation*}
\begin{equation*}
    \cdot e^{-\sum_i \frac{(\xi_i \sqrt{\alpha})^2 + (\Tilde{\xi}_i \sqrt{\alpha})^2}{2 \alpha}} \prod_i d (\xi_i \sqrt{\alpha}) d (\Tilde{\xi}_i \sqrt{\alpha}).
\end{equation*}

Again, we redefine $\xi_i \sqrt{\alpha} \longrightarrow \xi_i$, $\Tilde{\xi}_i \sqrt{\alpha} \longrightarrow \Tilde{\xi}_i$:

\begin{equation*}
    \hat{\rho} = \frac{1}{(2 \pi \alpha)^N} \int Tr_b \left\{ e^{\sum_i \xi_i (c \hat{a}_i^\dagger + s \hat{b}_i^\dagger)} \ket{0_a 0_b} \bra{0_a 0_b} e^{\sum_i \Tilde{\xi}_i (c \hat{a}_i + s \hat{b}_i)} \right\}  e^{-\sum_i \frac{\xi_i^2 + \Tilde{\xi}_i^2}{2 \alpha_i}} \prod_i d \xi_i d \Tilde{\xi}_i.
\end{equation*}

We compute partial trace over loss modes:

\begin{equation*}
    \hat{\rho} = \frac{1}{(2 \pi \alpha)^N } \int e^{\sum_i \xi_i c \hat{a}_i^\dagger } \ket{0} \bra{0} e^{\sum_i \Tilde{\xi}_i c \hat{a}_i} e^{-\sum_i \frac{\xi_i^2 + \Tilde{\xi}_i^2}{2 \alpha} + \xi_i \Tilde{\xi}_i s^2} \prod_i d \xi_i d \Tilde{\xi}_i.
\end{equation*}

This expression now can be considered as taking an expected value over a $2N$-dimensional normal distribution, where all variable pairs $\xi_i, \Tilde{\xi}_i$ are independent. Every variable pair $\xi_i, \Tilde{\xi}_i$ has covariance matrix $\Sigma$, and we can write this expression in the following way:

\begin{equation}
    \hat{\rho} = \frac{(det \Sigma)^{N/2}}{\alpha^N} \cdot \mathbb{E}_{\prod_i \mathbb{N}(0, \Sigma)} \left[ e^{\sum_i \xi_i c \hat{a}_i^\dagger } \ket{0} \bra{0} e^{\sum_i \Tilde{\xi_i} c \hat{a}_i} \right].
\end{equation}

For each variable pair $\xi_i, \Tilde{\xi}_i$ we now choose $\xi_{0i}, \chi_i, \Tilde{\chi}_i$ in a way that is described above. Then,

\begin{equation}
    \hat{\rho} = \frac{(det \Sigma)^{N/2}}{\alpha^N} \cdot \mathbb{E}_{\prod_i \mathbb{N}(0, \Gamma)} \left[ e^{\sum_i (\xi_{0i} + \chi_i) c \hat{a}_i^\dagger } \ket{0} \bra{0} e^{\sum_i (\xi_{0i} + \Tilde{\chi}_i) c \hat{a}_i} \right].
\end{equation}

We now consider the Taylor series expansion (up to the second order) of the expression in the square brackets, which we will denote $\hat{\mu}$:

\begin{equation*}
    \hat{\mu} = e^{\sum_i \chi_i c \hat{a}_i^\dagger} e^{\sum_i \xi_{0i} c \hat{a}_i^\dagger} \ket{0} \bra{0} e^{\sum_i \xi_{0i} c \hat{a}_i} e^{\sum_i \Tilde{\chi_i} c \hat{a}_i} = 
\end{equation*}
\begin{equation*}
    = \prod_i \left(1 + \chi_i c \hat{a}_i^\dagger + \frac{(\chi_i c \hat{a}_i^\dagger)^2}{2}\right) e^{\sum_i \xi_{0i} c \hat{a}_i^\dagger} \ket{0} \bra{0} e^{\sum_i \xi_{0i} c \hat{a}_i} \prod_i \left(1 + \Tilde{\chi_i} c \hat{a}_i + \frac{(\Tilde{\chi_i} c \hat{a}_i)^2}{2}\right).
\end{equation*}

The creation operators $\hat{a}_i^\dagger$ that act on the input modes can be written in terms of the operators $\hat{d}_i^\dagger$ that act on the output modes:

\begin{equation*}
    \hat{\mu} = \prod_i \left(1 + \chi_i c \sum_j U_{ij} \hat{d}_j^\dagger + \frac{(\chi_i c \sum_j U_{ij} \hat{d}_j^\dagger)^2}{2}\right) e^{\sum_{ij} \xi_{0i} c U_{ij}\hat{d}_j^\dagger} \ket{0} \cdot
\end{equation*}
\begin{equation*}    
\cdot \bra{0} e^{\sum_{ij} \xi_{0i} c U_{ij}^* \hat{d}_j} \prod_i \left(1 + \Tilde{\chi_i} c \sum_j U_{ij}^* \hat{d}_j + \frac{(\Tilde{\chi_i} c \sum_j U_{ij}^* \hat{d}_j)^2}{2}\right).
\end{equation*}

We will denote

\begin{equation}
    \hat{\nu}(\Vec{\xi}_0 c) = e^{\sum_{ij} \xi_{0i} c U_{ij}\hat{d}_j^\dagger} \ket{0} \bra{0} e^{\sum_{ij} \xi_{0i} c U_{ij}^* \hat{d}_j}.
\end{equation}

We can expand the brackets in the expression for $\hat{\mu}$, leaving the terms up to the second order:

\begin{equation*}    
\prod_i \left(1 + \chi_i c \sum_j U_{ij} \hat{d}^\dagger_j  + \frac{(\chi_i c \sum_j U_{ij} \hat{d}^\dagger_j )^2}{2}\right) = 1 + \sum_j \hat{d}^\dagger_j \sum_i c \chi_i U_{ij} + 
\end{equation*}
\begin{equation*}
    + \sum_{jk} \hat{d}^\dagger_j \hat{d}^\dagger_k \left( \frac{1}{2} \sum_i c^2 \chi^2_i U_{ij} U_{ik} + \sum_{i \neq l} c \chi_i c_l \chi_l U_{ij} U_{lk} \right).
\end{equation*}

\begin{equation*}    
\prod_i \left(1 + \Tilde{\chi_i} c \sum_j U_{ij}^* \hat{d}_j + \frac{(\Tilde{\chi_i} c \sum_j U_{ij}^* \hat{d}_j)^2}{2}\right) = 1 + \sum_j \hat{d}_j \sum_i c \Tilde{\chi_i}  U_{ij}^* + 
\end{equation*}
\begin{equation*}
    + \sum_{jk} \hat{d}_j \hat{d_k} \left( \frac{1}{2} \sum_i c^2 \Tilde{\chi}^2_i U_{ij}^* U_{ik}^* + \sum_{i \neq l} c \Tilde{\chi_i}  c_l \Tilde{\chi_l} U_{ij}^* U_{lk}^* \right).
\end{equation*}

When we take the product of these two expressions, most of the resulting terms will have zero expected value because of the properties of the normal distribution. Then

\begin{equation*}
    \hat{\mu} = \hat{\nu}(\Vec{\xi}_0 c) + \frac{1}{2} \sum_i \chi_i^2  c^2 \sum_{jk} U_{ij} U_{ik}  \cdot  \hat{d}^\dagger_j \hat{d}^\dagger_k \hat{\nu}(\Vec{\xi}_0 c) + \frac{1}{2} \sum_i  \Tilde{\chi_i}^2 c^2 \sum_{jk} U_{ij}^* U_{ik}^*  \cdot  \hat{\nu}(\Vec{\xi}_0 c) \hat{d}_j \hat{d}_k +
\end{equation*}
\begin{equation*}
    + \sum_i \chi_i \Tilde{\chi_i} c^2 \sum_{jk} U_{ij} U_{ik}^* \cdot \hat{d}^\dagger_j \hat{\nu}(\Vec{\xi}_0 c) \hat{d}_k + 
\end{equation*}
\begin{equation*}
    + \frac{1}{4} \sum_{ij} \chi_i^2 \Tilde{\chi_j}^2 c^4 \sum_{klmn} U_{ik} U_{il} U_{jm}^* U_{jn}^* \cdot \hat{d}^\dagger_k \hat{d}^\dagger_l \hat{\nu}(\Vec{\xi}_0 c) \hat{d}_m \hat{d}_n +
\end{equation*}
\begin{equation*}
    + \sum_{i \neq j} \chi_i \chi_j \Tilde{\chi_i} \Tilde{\chi_j} c^4 \sum_{klmn} U_{ik} U_{jl} U_{im}^* U_{jn}^* \cdot \hat{d}^\dagger_k \hat{d}^\dagger_l \hat{\nu}(\Vec{\xi}_0 c) \hat{d}_m \hat{d}_n
\end{equation*}

The integrals over $\chi_i$, $\Tilde{\chi}_i$ result in specific moments of the distribution, and the integral over $\xi_{0i}$ can be calculated using Monte-Carlo methods. The final expression is

\begin{equation*}
    Tr\left\{\hat{\rho}_{out} \hat{\Vec{n}}\right\} = \frac{(\det \Sigma)^{N/2}}{\alpha^N (2 \pi)^{3N/2} (\det \Gamma)^{N/2}} \cdot \int d\Vec{\xi}_0 \Bigg[ Tr\left\{\hat{\nu}(\Vec{\xi}_0 c) \hat{\Vec{n} }\right\} +
\end{equation*}
\begin{equation*}
    + \frac{1}{2} \overline{\chi^2}  c^2 \sum_{ijk} U_{ij} U_{ik}  \cdot Tr\left\{\hat{d}^\dagger_j \hat{d}^\dagger_k \hat{\nu}(\Vec{\xi}_0 c) \hat{\Vec{n}}\right\} + \frac{1}{2} \overline{\Tilde{\chi}^2}  c^2 \sum_{ijk} U_{ij}^* U_{ik}^*  \cdot Tr\left\{ \hat{\nu}(\Vec{\xi}_0 c) \hat{d}_j \hat{d}_k \hat{\Vec{n}}\right\} + 
\end{equation*}
\begin{equation*}
    + \overline{\chi \Tilde{\chi}}  c^2 \sum_{ijk} U_{ij} U_{ik}^* \cdot Tr\left\{\hat{d}^\dagger_j  \hat{\nu}(\Vec{\xi}_0 c) \hat{d}_k \hat{\Vec{n}}\right\} + 
\end{equation*}
\begin{equation*}
    + \frac{1}{4} c^4 \sum_{ij} \left( \left( \overline{\chi^2} \right)^2 + 2 \delta_{ij}  \left( \overline{\chi \Tilde{\chi}} \right)^2 \right) \sum_{klmn} U_{ik} U_{il} U_{jm}^* U_{jn}^* \cdot Tr\left\{ \hat{d}^\dagger_k \hat{d}^\dagger_l \hat{\nu}(\Vec{\xi}_0 c) \hat{d}_m \hat{d}_n \hat{\Vec{n}}\right\} + 
\end{equation*}
\begin{equation*}
    + \left( \overline{\chi\Tilde{\chi}} \right)^2 c^4 \sum_{i \neq j} \sum_{klmn} U_{ik} U_{jl} U_{im}^* U_{jn}^* \cdot Tr\left\{ \hat{d}^\dagger_k \hat{d}^\dagger_l \hat{\nu}(\Vec{\xi}_0 c) \hat{d}_m \hat{d}_n \hat{\Vec{n}}\right\} \Bigg],
\end{equation*}
where by Wick's probability theorem $\overline{\chi_i^2 \Tilde{\chi}_j^2} = \overline{\chi_i^2} \cdot \overline{\Tilde{\chi}_j^2} + 
\overline{\chi_i \Tilde{\chi}_j}\cdot \overline{\chi_i \Tilde{\chi}_j} + \overline{\chi_i \Tilde{\chi}_j}\cdot \overline{\chi_i \Tilde{\chi}_j} =  \left( \overline{\chi^2} \right)^2 + 2 \delta_{ij} \left( \overline{\chi \Tilde{\chi}} \right)^2$.

\subsection{Calculating traces}

In order to calculate  $Tr\left\{\hat{\rho}_{out} \hat{\Vec{n} }\right\}$, we need to be able to calculate expression like $Tr\left\{\hat{\nu}(\Vec{x}) \hat{\Vec{n} }\right\}$, $Tr\left\{ \hat{d}^\dagger_j \hat{d}^\dagger_k \hat{\nu}(\Vec{x}) \hat{\Vec{n} }\right\}$, $Tr\left\{ \hat{\nu}(\Vec{x}) \hat{d}_j \hat{d}_k  \hat{\Vec{n} }\right\}$, $Tr\left\{ \hat{d}^\dagger_j \hat{\nu}(\Vec{x}) \hat{d}_k  \hat{\Vec{n} }\right\}$, etc, for different $\Vec{x}$. The first one can be calculated fairly easily:

\begin{equation*}
    Tr\left\{\hat{\nu}(\Vec{x}) \hat{\Vec{n} }\right\} = Tr\left\{ e^{\sum_{ij} x_i U_{ij}\hat{d}_j^\dagger} \ket{0} \bra{0} e^{\sum_{ij} x_i U_{ij}^* \hat{d}_j} \ket{\Vec{n} } \bra{\Vec{n} }  \right\} = 
\end{equation*}
\begin{equation*}
     =  \bra{0} e^{\sum_{ij} x_i U_{ij}^* \hat{d}_j} \ket{\Vec{n} } \bra{\Vec{n} } e^{\sum_{ij} x_i c U_{ij}\hat{d}_j^\dagger} \ket{0} = 
\end{equation*}
\begin{equation*}
     = \prod_j \bra{0} e^{\sum_{i} x_i U_{ij}^* \hat{d}_j} \ket{n_j} \bra{n_j} e^{\sum_{i} x_i U_{ij}\hat{d}_j^\dagger} \ket{0} = 
\end{equation*}
\begin{equation*}
     = \prod_j \bra{0} \frac{\left(\sum_{i} x_i U_{ij}^* \hat{d}_j \right)^{n_j}}{n_j!} \ket{n_j} \bra{n_j} \frac{\left(\sum_{i} x_i U_{ij}\hat{d}_j^\dagger\right)^{n_j}}{n_j!} \ket{0} = 
\end{equation*}
\begin{equation*}
     = \prod_j \left[ \frac{\left(\sum_{i} x_i U_{ij}^* \right)^{n_j}}{\sqrt{n_j!}} \right] \cdot \left[ \frac{\left(\sum_{i} x_i U_{ij} \right)^{n_j}}{\sqrt{n_j!}} \right] = 
\end{equation*}
\begin{equation*}
     = \prod_j \frac{1}{n_j!} \left| \sum_{i} x_i U_{ij} \right|^{2n_j}.
\end{equation*}

Now suppose we need to calculate $Tr\left\{ (\hat{d}^\dagger_1)^{q_1} ... (\hat{d}^\dagger_N)^{q_N} \hat{\nu}(\Vec{x}) (\hat{d}_1)^{q_1} ... (\hat{d}_N)^{q_N}  \hat{\Vec{n} }\right\}$. First, we note that 

\begin{equation*}
    Tr\left\{ (\hat{d}^\dagger_1)^{q_1} ... (\hat{d}^\dagger_N)^{q_N} \hat{\nu}(\Vec{x}) (\hat{d}_1)^{p_1} ... (\hat{d}_N)^{p_N}  \hat{\Vec{n} }\right\} = Tr\left\{ \hat{\nu}(\Vec{x}) (\hat{d}_1)^{p_1} ... (\hat{d}_N)^{p_N}  \hat{\Vec{n} } (\hat{d}^\dagger_1)^{q_1} ... (\hat{d}^\dagger_N)^{q_N} \right\} = 
\end{equation*}
\begin{equation*}
     = Tr\left\{ \hat{\nu}(\Vec{x}) \ket{\Vec{n}-\Vec{p}} \bra{\Vec{n}-\Vec{q}} \right\} \cdot \prod_j \sqrt{n_j (n_j-1) ... (n_j-p_j+1) n_j (n_j-1) ... (n_j-q_j+1)} =
\end{equation*}
\begin{equation*}
     = Tr\left\{ \hat{\nu}(\Vec{x}) \ket{\Vec{n}-\Vec{p}} \bra{\Vec{n}-\Vec{q}} \right\} \cdot \prod_j \sqrt{\frac{n_j!}{(n_j-p_j)!} \frac{n_j!}{(n_j-q_j)!}},
\end{equation*}
where by e.g. $\ket{\Vec{n}-\Vec{p}}$ we mean $\bigotimes_i \ket{n_i-p_i}$.

\begin{equation*}
     = Tr\left\{ \hat{\nu}(\Vec{x}) \ket{\Vec{n}-\Vec{p}} \bra{\Vec{n}-\Vec{q}} \right\} = \prod_j \bra{0} e^{\sum_{i} x_i U_{ij}^* \hat{d}_j} \ket{n_j-p_j} \bra{n_j-q_j} e^{\sum_{i} x_i U_{ij}\hat{d}_j^\dagger} \ket{0} = 
\end{equation*}
\begin{equation*}
     = \prod_j \left[ \frac{\left(\sum_{i} x_i U_{ij}^* \right)^{n_j-p_j}}{\sqrt{(n_j-p_j)!}} \right] \cdot \left[ \frac{\left(\sum_{i} x_i U_{ij} \right)^{n_j-q_j}}{\sqrt{(n_j-q_j)!}} \right] = 
\end{equation*}
\begin{equation*}
     = \prod_j \frac{1}{\sqrt{(n_j-p_j)! (n_j-q_j)!}} \cdot \frac{\left|\sum_{i} x_i U_{ij} \right|^{2 n_j}}{\left(\sum_{i} x_i U_{ij}^* \right)^{p_j} \left(\sum_{i} x_i U_{ij} \right)^{q_j}} = 
\end{equation*}
\begin{equation*}
     = \prod_j \frac{\left|\sum_{i} x_i U_{ij} \right|^{2 n_j}}{\left(\sum_{i} x_i U_{ij}^* \right)^{p_j} \left(\sum_{i} x_i U_{ij} \right)^{q_j}} \cdot \frac{\sqrt{\frac{n_j!}{(n_j-p_j)!} \frac{n_j!}{(n_j-q_j)!}}}{n_j!} =
\end{equation*}
\begin{equation*}
     = Tr\left\{\hat{\nu}(\Vec{x}) \hat{\Vec{n} }\right\} \prod_j \sqrt{\frac{n_j!}{(n_j-p_j)!} \frac{n_j!}{(n_j-q_j)!}} \frac{1}{\left(\sum_{i} x_i U_{ij}^* \right)^{p_j} \left(\sum_{i} x_i U_{ij} \right)^{q_j}}.
\end{equation*}

Finally, we can write

\begin{equation*}
    Tr\left\{ (\hat{d}^\dagger_1)^{q_1} ... (\hat{d}^\dagger_N)^{q_N} \hat{\nu}(\Vec{x}) (\hat{d}_1)^{p_1} ... (\hat{d}_N)^{p_N}  \hat{\Vec{n} }\right\} = 
\end{equation*}
\begin{equation*}
    = Tr\left\{\hat{\nu}(\Vec{x}) \hat{\Vec{n} }\right\} \prod_j \frac{n_j!}{(n_j-p_j)!} \frac{n_j!}{(n_j-q_j)!} \frac{1}{\left(\sum_{i} x_i U_{ij}^* \right)^{p_j} \left(\sum_{i} x_i U_{ij} \right)^{q_j}}.
\end{equation*}

\section{Algorithm overview}

The goal of the algorithm is to calculate the probability of a state $\ket{\Vec{n}}$, given $\Vec{n}$, $\alpha$, $c$, $s$ and $U$. We assume that the Taylor series expansion is done up to the desired order before computation starts. The integrals over $\chi_i$ and $\Tilde{\chi}_i$ should also be computed (it can be done analytically via Wick's probability theorem).

We start by calculating two-variable covariance matrix $\Sigma$ using $\alpha$ and $s$. We now select $\Gamma$ in the way specified above such that it minimizes the series expansion parameter $\varepsilon$. In order to compute the integrals over $\xi_{0i}$, we sample $\xi_{0i}$ for each $i$ from a normal distribution $\mathbb{N}(0, \overline{\xi_0^2})$. 

We now compute $Tr\left\{\hat{\mu} \hat{\Vec{n} }\right\}$, which by linearity consists in computing traces of the form described above; for each sample $\Vec{\xi}_0$ we need only a polynomial number of operations. 

Finally, we take an average over our samples and multiply by the necessary constant terms.

\section{Taylor series convergence for actual experimental conditions}

We have discussed above the fact that the role of the "perturbation parameter"\ in the series expansion is played by $c^2 \cdot \min(\overline{\chi^2}, |h|)$, which we can choose to be equal to $\varepsilon=\frac{1}{2}\frac{c^2}{1/\alpha + s^2}$. This parameter depends on the experimental conditions (i.e. the squeezing parameter of the input state $\alpha$ and loss level $s^2$). The smaller this parameter is, the faster the series will converge. Thus, the best conditions for this algorithm are achieved when the loss level $s^2$ is high and the squeezing parameter $\alpha$ is low. Let us consider actual experimental implementation of the gaussian boson sampling problem, and estimate how small this parameter is in those conditions.

Let's consider the relation between $\alpha$ and the average amount of photons per state $\langle n \rangle$. If the squeezing parameter is $\zeta = r e^{i \varphi}$, then $\alpha = \tanh{r}$, while $\langle n \rangle = \sinh^2{r}$.

% https://www.science.org/doi/full/10.1126/science.abe8770
% average collection efficiency = 0.628
% 25 TMSS = 50 SMSS; 43 photons average => <n> = 43/50
% eps ~ 0.18

In a paper by Zhong et al. \cite{Zhong_2020} 25 PPKTP crystals were used to produce 25 two-mode squeezed states, which is equivalent to 50 single-mode squeezed states. The average number of photons registered by the detectors is 43. Thus, the average amount of photons per mode $\langle n \rangle$ is around $\frac{43}{50}$; $r=\arcsinh (\sqrt{\langle n \rangle}) \approx 0.855$, $\alpha = \tanh{r} \approx 0.694$. The average collection efficiency is said to be $c^2 = 0.628$. Then, $\varepsilon = \frac{1}{2}\frac{c^2}{\frac{1}{\alpha}+s^2} \approx 0.18$.

% Zhong_2021
% average transmission coefficient 0.48-0.54
% 50 SMSS; 70 photons average at max pump
% eps ~ 0.14

In another paper by Zhong et al. \cite{Zhong_2021}, the average amount of photons produced was increased to $70$ at maximum pump intensity. This corresponds 
to $\alpha \approx 0.76$. The overall transmission rate in the experiment is said in the paper to be $48\%$ and $54\%$ for different settings, so we take $s^2 \approx 0.5$. This yields $\varepsilon \approx 0.14$.

The conclusion that we draw is that even in large GBS experiments which are said to demonstrate quantum advantage the conditions are such that $\varepsilon$ is fairly small, and not many orders of the series need to be calculated to produce an approximation. 

\section{Implementation details}

\subsection{Contraction precomputation}

Let's consider the term

\begin{equation*}
    \frac{1}{2} \overline{\chi^2}  c^2 \sum_{ijk} U_{ij} U_{ik}  \cdot Tr\left\{\hat{d}^\dagger_j \hat{d}^\dagger_k \hat{\nu}(\Vec{\xi}_0 c) \hat{\Vec{n}}\right\}.
\end{equation*}

We can rewrite it as

\begin{equation*}
    \frac{1}{2} \overline{\chi^2}  c^2 \sum_{jk} Tr\left\{\hat{d}^\dagger_j \hat{d}^\dagger_k \hat{\nu}(\Vec{\xi}_0 c) \hat{\Vec{n}}\right\} \sum_{i} U_{ij} U_{ik} = \frac{1}{2} \overline{\chi^2}  c^2 \sum_{jk} Tr\left\{\hat{d}^\dagger_j \hat{d}^\dagger_k \hat{\nu}(\Vec{\xi}_0 c) \hat{\Vec{n}}\right\} T_{jk},
\end{equation*}
where $T_{jk} = \sum_{i} U_{ij} U_{ik}$ is a contraction of $U$ with itself. It depends only on $U$, and can be calculated before sampling $\Vec{\xi}_0$, which reduces the amount of operations required to calculate each probability sample from a $\Vec{\xi}_0$ sample.

\subsection{Factorial fractions precomputation}

In calculating traces of the form described above, we need to calculate factorial fractions of the form $\frac{m!}{(m-p)!} \equiv F^{m}_{p}$, where $0 \leq p \leq m$. Since the target state $\hat{\Vec{n}}$ is fixed, $m \leq \max(n_i)$.

\subsection{Reusing $\sum_{i} x_i U_{ij}$}

During calculation, while calculating each trace, we can calculate $\sum_{i} x_i U_{ij}$ only once for each $\Vec{\xi}_0$ sample and then reuse it, thus using less operations to calculate each trace. Let's denote $S_j = \sum_{i} x_i U_{ij}$; $\Vec{S}=U^T \Vec{x}$. Then, 

\begin{equation}
    Tr\left\{\hat{\nu}(\Vec{x}) \hat{\Vec{n} }\right\} = \prod_j \frac{1}{n_j!} \left| S_j \right|^{2n_j} 
\end{equation}
and

\begin{equation*}
    Tr\left\{ (\hat{d}^\dagger_1)^{q_1} ... (\hat{d}^\dagger_N)^{q_N} \hat{\nu}(\Vec{x}) (\hat{d}_1)^{p_1} ... (\hat{d}_N)^{p_N}  \hat{\Vec{n} }\right\} = 
\end{equation*}
\begin{equation*}
    = Tr\left\{\hat{\nu}(\Vec{x}) \hat{\Vec{n} }\right\} \prod_j \frac{n_j (n_j-1) ... (n_j-p_j+1) n_j (n_j-1) ... (n_j-q_j+1)}{\left(S_j^* \right)^{p_j} \left( S_j \right)^{q_j}}.
\end{equation*}

\section{Complexity analysis}

\subsection{Precomputation}

In this section we will analyze the computational complexity of precomputation. By precomputation we mean the calculations that need to be carried out only once before $\Vec{\xi}_0$ sampling and before calculating probability samples for each $\Vec{\xi}_0$. The multiplicative constant before the integral sign $\frac{(\det \Sigma)^{N/2}}{\alpha^N (2 \pi)^{3N/2} (\det \Gamma)^{N/2}}$ can be calculated with $O(N)$ multiplication operations. For each term in the resulting sum, we will define its order to be the number of variables $\chi$ and $\Tilde{\chi}$, or, equivalently, the power of the loss parameter $c$. Thus, the term 

\begin{equation*}
    \overline{\chi \Tilde{\chi}}  c^2 \sum_{ijk} U_{ij} U_{ik}^* \cdot Tr\left\{\hat{d}^\dagger_j  \hat{\nu}(\Vec{\xi}_0 c) \hat{d}_k \hat{\Vec{n}}\right\}
\end{equation*}
will be of the second order. Then, each term of the order $K$ will have a contraction of the form

\begin{equation}
    \sum_{j_1 ... j_K} U_{i_1 j_1} U_{i_2 j_2} ... U_{i_K j_K}
\end{equation}
where some of the $U_{ji}$ can be conjugated. This leaves at most $K+1$ different ways to conjugate the factors. Each contraction has $K$ free indices, and calculating the sum requires $N^K$ additions and $N^K(K-1)$ multiplications. The total number of additions is $N^{2K}$ and the number of multiplications is $N^{2K}(K-1)$, where $K$ is the maximum order we choose to calculate.

Calculating all $F^{m}_{p} \equiv \frac{m!}{(m-p)!}$ for $0 \leq p \leq m \leq \max(n_i)$ requires only around $\frac{\max(n_i)^2}{2}$ multiplications, since $\forall m ~~ F^m_0 = 1$, $F^m_1 = m$, $F^m_2 = m(m-1) = (m-1) F^m_1$, ..., $F^m_k = (m-k+1)F^m_{k-1}$.

\subsection{Probability sample computation}

Here we will analyze the computational complexity of calculating a single probability sample given $\Vec{\xi}_0$. We will assume that the terms are calculated up to some order $K$.

Calculating the trace 

\begin{equation*}
    Tr\left\{\hat{\nu}(\Vec{x}) \hat{\Vec{n} }\right\}     = \prod_j \frac{1}{n_j!} \left| \sum_{i} x_i U_{ij} \right|^{2n_j}.
\end{equation*}
requires one multiplication of an $N\times N$ matrix by a $N$-dimensional vector, $N$ exponentiation operations and $2N$ multiplication operations. This calculations needs to be done only once for each $\Vec{x}$. Calculating any other trace of the form 

\begin{equation*}
    Tr\left\{ (\hat{d}^\dagger_1)^{q_1} ... (\hat{d}^\dagger_N)^{q_N} \hat{\nu}(\Vec{x}) (\hat{d}_1)^{p_1} ... (\hat{d}_N)^{p_N}  \hat{\Vec{n} }\right\}
\end{equation*}
requires $2N$ exponentiation operations and $4N$ multiplication operations (since factorial fractions are precomputed).

The number of terms for a given order $K$ is $N^K$ times the number of different non-zero $K$-th order moments $\overline{\chi_{i_1} \dotsm \chi_{i_r} \Tilde{\chi}_{i_{r+1}} \dotsm \Tilde{\chi}_{i_K}}$. The exact amount is hard to calculate, but the total number of moments (including those that are zero) is $(K+1)N^K$. Thus, the maximum amount of terms required to is $(K+1)N^{2K}$. 

Since the amount of operations required to calculate each term is $O(N)$, the total computational complexity of calculating a probability sample for a given $\Vec{\xi}_0$ is $O \left( K \cdot N^{2K} \right)$

\section{Results}

Below are the results of probability calculation for $N=5$ for different output states. The calculated probabilities are compared to exact solutions. The parameters are: $\alpha = 0.9$, $c=s=\sqrt{2}$. Number of samples is $4096$.

\begin{figure}[H]
\centering
\includegraphics[scale=0.5]{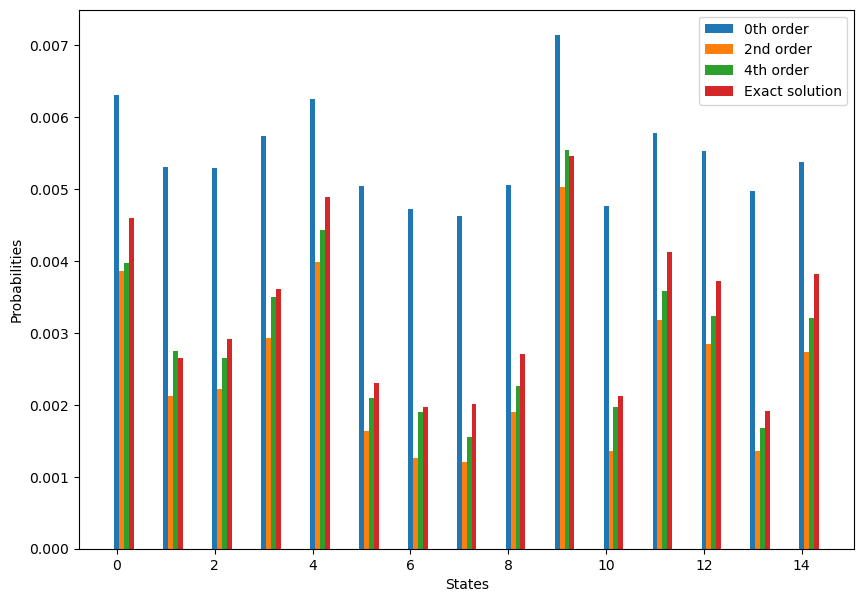}
\caption{Probability calculation for $5$ modes for different $2$-photon output states.}
\end{figure}

\begin{figure}[H]
\centering
\includegraphics[scale=0.5]{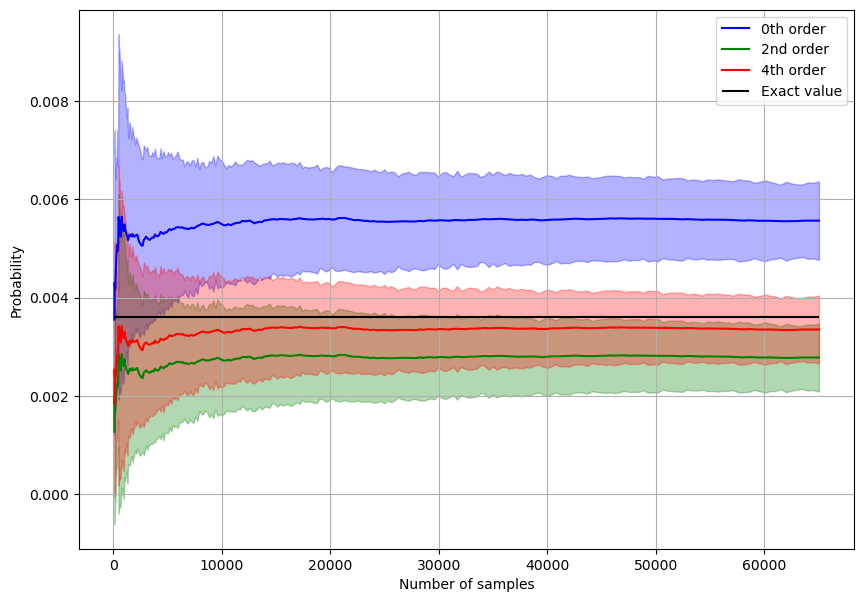}
\caption{Graph of the average probability and the standard deviation calculated up to different orders for different numbers of samples. The state for this graph is $2$-photon.}
\end{figure}

These results show that for calculating a single output state probability accurately the number of samples needs to be on the order of $10^4$. Below are the results of using fewer samples per state, but instead of comparing individual probabilities we look at cosine similarity between exact and approximated probability distributions over all 2-photon states.

\begin{figure}[H]
\centering
\includegraphics[scale=0.5]{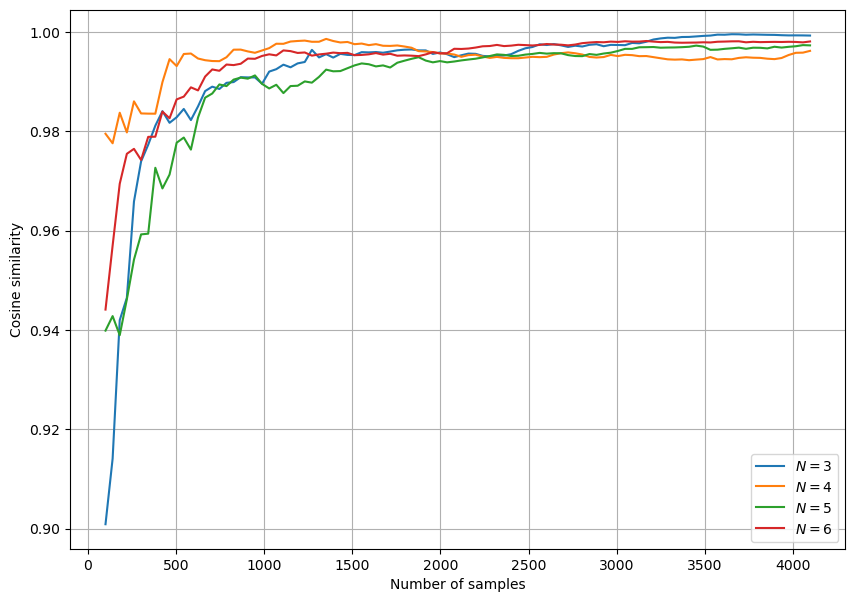}
\caption{Convergence of the cosine similarity between estimated probability distribution over the set of all 2-photon states and ground truth for different $N$.}
\end{figure}

The above graph suggests that the number of samples per state needed to approximate the distribution doesn't depend much on $N$. It is computationally hard to check this when comparing to the exact solution, but if we assume that the cosine similarity converges to a value close to $1$, we can estimate how quickly it converges. Below we look at the cosine similarity between a distribution calculated with $K$ samples per state and a distribution calculated with $K+10$ samples per state for different K. Figure 4 suggests more strongly that the number of samples per state required for accurate approximation is not really influenced by $N$

\begin{figure}[H]
\centering
\includegraphics[scale=0.5]{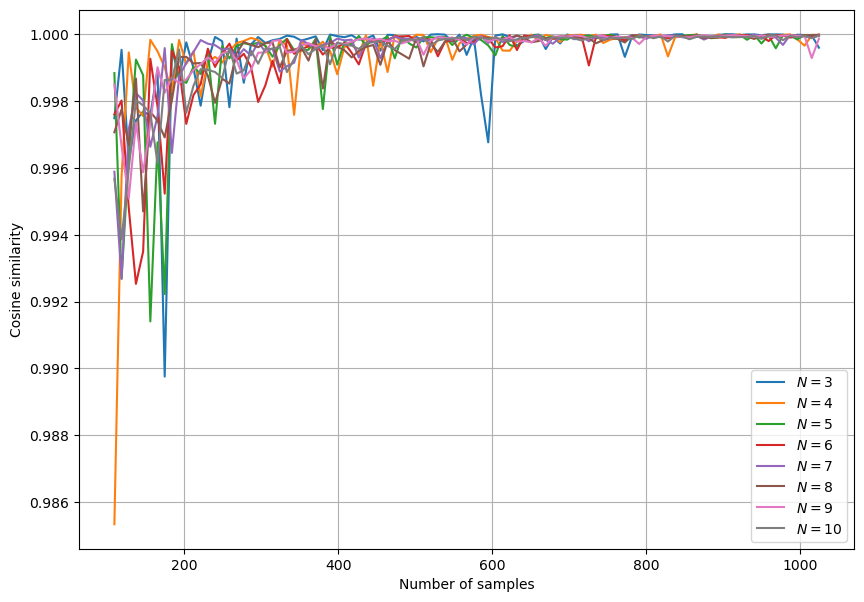}
\caption{Cosine similarity between probability distribution over the set of all 2-photon states after $K$ samples and after $K+10$ samples for different $N$.}
\end{figure}

% TODO
% \begin{figure}[H]
% \centering
% \includegraphics[scale=0.5]{CS_exact_C2.png}
% \caption{Convergence of the cosine similarity between estimated probability distribution over the set of all 2-photon states and ground truth for different $c^2$ (resulting in different $\varepsilon$).}
% \end{figure}

Below are benchmark results that show average precomputation time, which depends only on $N$, and time per sample, which depends on $N$ and the amount of photons $M$ in the target state. 

\begin{figure}[H]
\centering
\includegraphics[scale=0.5]{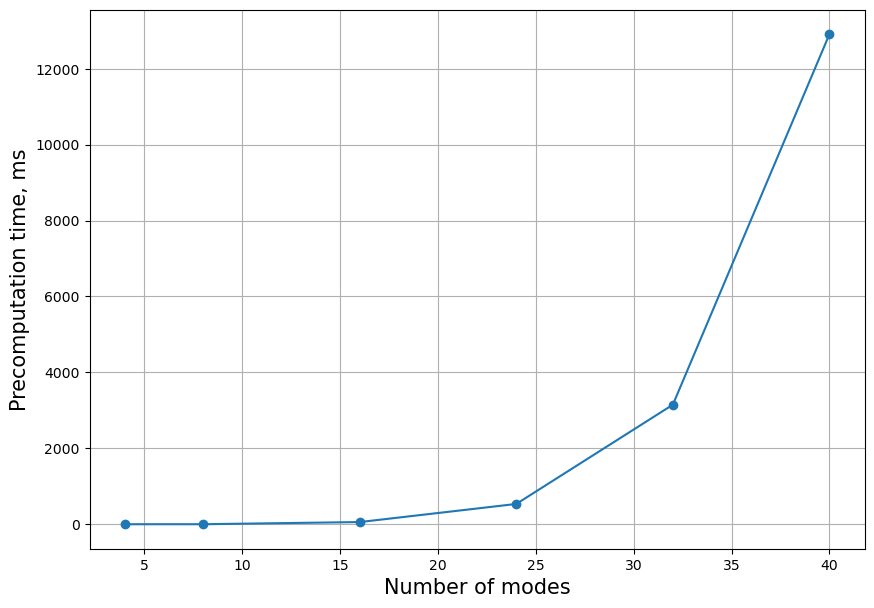}
\caption{Precomputation time on an Intel i5 CPU in ms versus the number of modes.}
\end{figure}

\begin{figure}[H]
\centering
\includegraphics[scale=0.5]{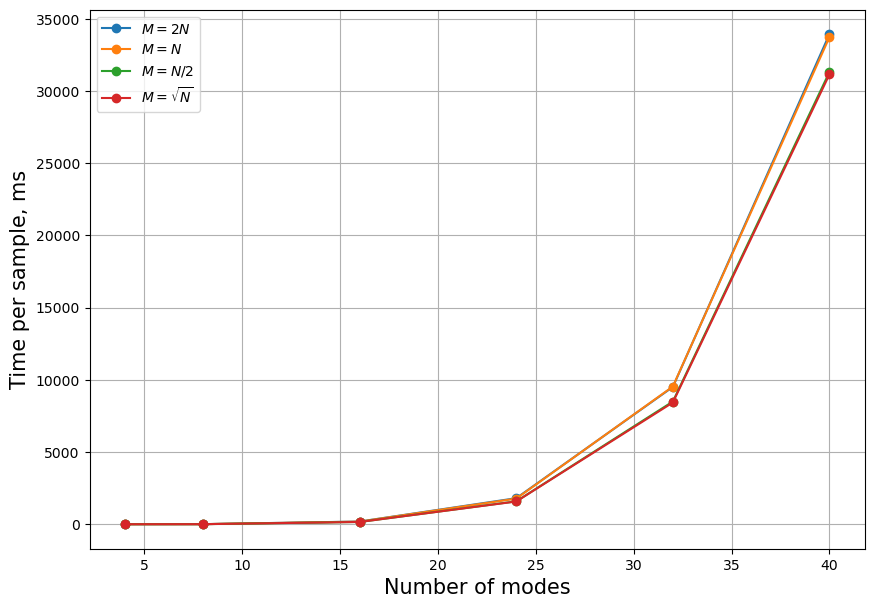}
\caption{Average time per sample on an Intel i5 CPU versus the number of modes for states with different photon numbers.}
\end{figure}

These results show that even $N=40$ mode GBS can be simulated on an average laptop using this algorithm.

\section{Conclusion}

In this paper we have presented a new algorithm for the approximate calculation of the probability of observing a given output state in Gaussian boson sampling instance. We have discussed various implementation details that help to reduce the number of operations needed to calculate each probability sample. We also analyze the total computational complexity both of the calculations that need to be carried out once for each specific problem and of computing each probability sample.

This algorithm relies on the Taylor series expansion where the "perturbation"\ parameter is dependent on the problem conditions. The algorithm consists in calculating the terms of this Taylor series up to some finite order. For a fixed maximum order, the computational complexity of the algorithm is polynomial in $N$.

We have demonstrated that increasing the maximum order does increase the accuracy of the answer. We have also measured precomputation and sampling time for a regular CPU, showing that even large instances of Gaussian boson sampling ($N \approx 40$) can be solved in reasonable time. 

We have considered recent GBS experiments and estimated the parameters of the problem for those conditions. We believe that in those conditions our algorithm doesn't require many orders of the Taylor series to be calculated for approximating a probability of an output state.

\printbibliography

@misc{https://doi.org/10.48550/arxiv.2106.01445,
  doi = {10.48550/ARXIV.2106.01445},
  url = {https://arxiv.org/abs/2106.01445},
  author = {Popova, A. S. and Rubtsov, A. N.},
  title = {Cracking the Quantum Advantage threshold for Gaussian Boson Sampling},
  publisher = {arXiv},
  year = {2021},
  copyright = {arXiv.org perpetual, non-exclusive license}
}

@article{Garc_a_Patr_n_2019,
	doi = {10.22331/q-2019-08-05-169},
	url = {https://doi.org/10.22331%2Fq-2019-08-05-169},
	year = 2019,
	month = {aug},
	publisher = {Verein zur Forderung des Open Access Publizierens in den Quantenwissenschaften},
	volume = {3},
	pages = {169},
	author = {Raúl García-Patrón and Jelmer J. Renema and Valery Shchesnovich},
	title = {Simulating boson sampling in lossy architectures},
	journal = {Quantum}
}

@article{Oh_2021,
	doi = {10.1103/physreva.104.022407},
	url = {https://doi.org/10.1103%2Fphysreva.104.022407},
	year = 2021,
	month = {aug},
	publisher = {American Physical Society ({APS})},
	volume = {104},
	number = {2},
	author = {Changhun Oh and Kyungjoo Noh and Bill Fefferman and Liang Jiang},
	title = {Classical simulation of lossy boson sampling using matrix product operators},
	journal = {Physical Review A}
}

@incollection{Gard_2015,
	doi = {10.1142/9789814678704_0008},
	url = {https://doi.org/10.1142%2F9789814678704_0008},
	year = 2015,
	month = {jun},
	publisher = {{WORLD} {SCIENTIFIC}
},
	pages = {167--192},
	author = {Bryan T. Gard and Keith R. Motes and Jonathan P. Olson and Peter P. Rohde and Jonathan P. Dowling},
	title = {An Introduction to Boson-Sampling},
	booktitle = {From Atomic to Mesoscale}
}

@article{Lund_2014,
	doi = {10.1103/physrevlett.113.100502},
	url = {https://doi.org/10.1103%2Fphysrevlett.113.100502},
	year = 2014,
	month = {sep},
	publisher = {American Physical Society ({APS})},
	volume = {113},
	number = {10},
	author = {A.{\hspace{0.167em}
}P. Lund and A. Laing and S. Rahimi-Keshari and T. Rudolph and J.{\hspace{0.167em}}L. O'Brien and T.{\hspace{0.167em}}C. Ralph},
	title = {Boson Sampling from a Gaussian State},
	journal = {Physical Review Letters}
}

@article{Shor_1997,
	doi = {10.1137/s0097539795293172},
	url = {https://doi.org/10.1137%2Fs0097539795293172},
	year = 1997,
	month = {oct},
	publisher = {Society for Industrial {\&} Applied Mathematics ({SIAM})},
	volume = {26},
	number = {5},
	pages = {1484--1509},
	author = {Peter W Shor},
	title = {Polynomial-Time Algorithms for Prime Factorization and Discrete Logarithms on a Quantum Computer},
	journal = {{SIAM} Journal on Computing}
}

@article{Qi_2020,
	doi = {10.1103/physrevlett.124.100502},
	url = {https://doi.org/10.1103%2Fphysrevlett.124.100502},
	year = 2020,
	month = {mar},
	publisher = {American Physical Society ({APS})},
	volume = {124},
	number = {10},
	author = {Haoyu Qi and Daniel J. Brod and Nicol{\'{a}
}s Quesada and Raúl García-Patrón},
	title = {Regimes of Classical Simulability for Noisy Gaussian Boson Sampling},
	journal = {Physical Review Letters}
}

@article{Aaronson_2016,
	doi = {10.1103/physreva.93.012335},
	url = {https://doi.org/10.1103%2Fphysreva.93.012335},
	year = 2016,
	month = {jan},
	publisher = {American Physical Society ({APS})},
	volume = {93},
	number = {1},
	author = {Scott Aaronson and Daniel J. Brod},
	title = {{BosonSampling} with lost photons},
	journal = {Physical Review A}
}

@article{Zhong_2020,
	doi = {10.1126/science.abe8770},
	url = {https://doi.org/10.1126%2Fscience.abe8770},
	year = 2020,
	month = {dec},
	publisher = {American Association for the Advancement of Science ({AAAS})},
	volume = {370},
	number = {6523},
	pages = {1460--1463},
	author = {Han-Sen Zhong and Hui Wang and Yu-Hao Deng and Ming-Cheng Chen and Li-Chao Peng and Yi-Han Luo and Jian Qin and Dian Wu and Xing Ding and Yi Hu and Peng Hu and Xiao-Yan Yang and Wei-Jun Zhang and Hao Li and Yuxuan Li and Xiao Jiang and Lin Gan and Guangwen Yang and Lixing You and Zhen Wang and Li Li and Nai-Le Liu and Chao-Yang Lu and Jian-Wei Pan},
	title = {Quantum computational advantage using photons},
	journal = {Science}
}

@article{PhysRevLett.113.100502,
  title = {Boson Sampling from a Gaussian State},
  author = {Lund, A. P. and Laing, A. and Rahimi-Keshari, S. and Rudolph, T. and O'Brien, J. L. and Ralph, T. C.},
  journal = {Phys. Rev. Lett.},
  volume = {113},
  issue = {10},
  pages = {100502},
  numpages = {5},
  year = {2014},
  month = {Sep},
  publisher = {American Physical Society},
  doi = {10.1103/PhysRevLett.113.100502},
  url = {https://link.aps.org/doi/10.1103/PhysRevLett.113.100502}
}

@article{SBS_exp,
doi = 10.1126/sciadv.1400255,
author = {Marco Bentivegna  and Nicolò Spagnolo  and Chiara Vitelli  and Fulvio Flamini  and Niko Viggianiello  and Ludovico Latmiral  and Paolo Mataloni  and Daniel J. Brod  and Ernesto F. Galvão  and Andrea Crespi  and Roberta Ramponi  and Roberto Osellame  and Fabio Sciarrino },
title = {Experimental scattershot boson sampling},
journal = {Science Advances},
volume = {1},
number = {3},
pages = {e1400255},
year = {2015},
doi = {10.1126/sciadv.1400255},
URL = {https://www.science.org/doi/abs/10.1126/sciadv.1400255},
eprint = {https://www.science.org/doi/pdf/10.1126/sciadv.1400255}}

@article{Hamilton_2017,
	doi = {10.1103/physrevlett.119.170501},
	url = {https://doi.org/10.1103%2Fphysrevlett.119.170501},
	year = 2017,
	month = {oct},
	publisher = {American Physical Society ({APS})},
	volume = {119},
	number = {17},
	author = {Craig S. Hamilton and Regina Kruse and Linda Sansoni and Sonja Barkhofen and Christine Silberhorn and Igor Jex},
	title = {Gaussian Boson Sampling},
	journal = {Physical Review Letters}
}

@article{Zhong_2019,
	doi = {10.1016/j.scib.2019.04.007},
    url = {https://doi.org/10.1016%2Fj.scib.2019.04.007},
	year = 2019,
	month = {apr},
	publisher = {Elsevier {BV}
},
	volume = {64},
	number = {8},
	pages = {511--515},
	author = {Han-Sen Zhong and Li-Chao Peng and Yuan Li and Yi Hu and Wei Li and Jian Qin and Dian Wu and Weijun Zhang and Hao Li and Lu Zhang and Zhen Wang and Lixing You and Xiao Jiang and Li Li and Nai-Le Liu and Jonathan P. Dowling and Chao-Yang Lu and Jian-Wei Pan},
	title = {Experimental Gaussian Boson sampling},
	journal = {Science Bulletin}
}

@article{PhysRevLett.3.77,
  title = {Calculation of Partition Functions},
  author = {Hubbard, J.},
  journal = {Phys. Rev. Lett.},
  volume = {3},
  issue = {2},
  pages = {77--78},
  numpages = {0},
  year = {1959},
  month = {Jul},
  publisher = {American Physical Society},
  doi = {10.1103/PhysRevLett.3.77},
  url = {https://link.aps.org/doi/10.1103/PhysRevLett.3.77}
}

@ARTICLE{1957SPhD....2..416S,
       author = {{Stratonovich}, R.~L.},
        title = "{On a Method of Calculating Quantum Distribution Functions}",
      journal = {Soviet Physics Doklady},
         year = 1957,
        month = jul,
       volume = {2},
        pages = {416},
       adsurl = {https://ui.adsabs.harvard.edu/abs/1957SPhD....2..416S}
}

@article{Zhong_2021,
   title={Phase-Programmable Gaussian Boson Sampling Using Stimulated Squeezed Light},
   volume={127},
   ISSN={1079-7114},
   url={http://dx.doi.org/10.1103/PhysRevLett.127.180502},
   DOI={10.1103/physrevlett.127.180502},
   number={18},
   journal={Physical Review Letters},
   publisher={American Physical Society (APS)},
   author={Zhong, Han-Sen and Deng, Yu-Hao and Qin, Jian and Wang, Hui and Chen, Ming-Cheng and Peng, Li-Chao and Luo, Yi-Han and Wu, Dian and Gong, Si-Qiu and Su, Hao and Hu, Yi and Hu, Peng and Yang, Xiao-Yan and Zhang, Wei-Jun and Li, Hao and Li, Yuxuan and Jiang, Xiao and Gan, Lin and Yang, Guangwen and You, Lixing and Wang, Zhen and Li, Li and Liu, Nai-Le and Renema, Jelmer J. and Lu, Chao-Yang and Pan, Jian-Wei},
   year={2021},
   month=oct }

@misc{oh2023classical,
      title={Classical algorithm for simulating experimental Gaussian boson sampling}, 
      author={Changhun Oh and Minzhao Liu and Yuri Alexeev and Bill Fefferman and Liang Jiang},
      year={2023},
      eprint={2306.03709},
      archivePrefix={arXiv},
      primaryClass={quant-ph}
}

%\section{Appendix}

\end{document}